# Privacy, Robustness, and Fairness Trade-offs in Federated Intrusion Detection: Geometric Indistinguishability at the Aggregation Interface

Adrita Rahman Tory[1], ABM Shawkat Ali[1], Md Abu Layek[2], and Khondokar Fida Hasan[3]*

[1] Bangladesh University of Business and Technology (BUBT), Mirpur-2, Dhaka-1216, Bangladesh
[2] Jagannath University, Dhaka, Bangladesh
[3] University of New South Wales (UNSW), ACT 2601, Australia
fida.hasan@unsw.edu.au

**Abstract.** Federated learning enables privacy-conscious collaboration for network intrusion detection without centralizing sensitive traffic data, yet its deployment in operational environments must simultaneously satisfy three competing requirements: formal differential privacy guaranties, tolerance to Byzantine-adversarial participants, and reliable detection coverage across severely imbalanced attack categories. Existing literature treats these properties as independently composable, an assumption that this paper challenges both theoretically and empirically. In this paper, we study how these requirements interact in class-imbalanced federated NIDS and introduce geometric indistinguishability as a conceptual lens for a regime in which privacy-induced dispersion in client updates can make minority-class signals harder for robust aggregation to preserve. Using UNSW-NB15 as a case study, we evaluate DP-SGD combined with coordinate-wise median under label-flip and model-poisoning attacks, with threat coverage assessed across attack categories. Our results provide initial evidence that the joint use of privacy noise and robust aggregation can disproportionately degrade detection of rare attacks relative to majority classes. We also show that part of the observed collapse under strong privacy can arise from training miscalibration, while a residual performance floor may remain for ultra-rare categories even after $\epsilon$-dependent tuning. These findings motivate studying privacy, robustness, and rare-attack coverage jointly rather than as independently composable properties, and suggest that aggregation-aware modeling and sample-aware evaluation are promising directions for trustworthy federated NIDS.

**Keywords:** Federated Learning · Intrusion Detection · Differential Privacy · Byzantine Robustness · Minority Attack Detection · Network Security

* Corresponding author, *Email: fida.hasan@unsw.edu.au*

## 1 Introduction

Machine-learning-based Network Intrusion Detection Systems (NIDS) face a persistent challenge in recognising attack categories that vary enormously in prevalence. In operational environments, rare and novel attacks can be disproportionately important despite appearing only sparsely in observed traffic. At the same time, any single organization typically sees only a partial slice of the broader threat landscape, creating a strong incentive for collaborative learning across administrative boundaries [9, 12, 14, 15, 18]. In practice, however, network traffic metadata may constitute personal data under regulatory frameworks such as the GDPR, which limits the feasibility of centralizing raw data [7].

Federated Learning (FL), a distributed paradigm where participants share model parameter updates rather than raw data [1], enables such collaboration while respecting data protection constraints. However, FL provides no formal privacy guarantees by default; gradient inversion attacks can reconstruct sensitive training inputs from shared updates [1, 8]. Consequently, Differentially Private Stochastic Gradient Descent (DP-SGD) [1, 8, 19] is widely adopted to bound this leakage, providing $(\varepsilon, \delta)$-privacy guarantees by clipping per-sample gradients and injecting calibrated Gaussian noise [6,8], albeit at utility cost.

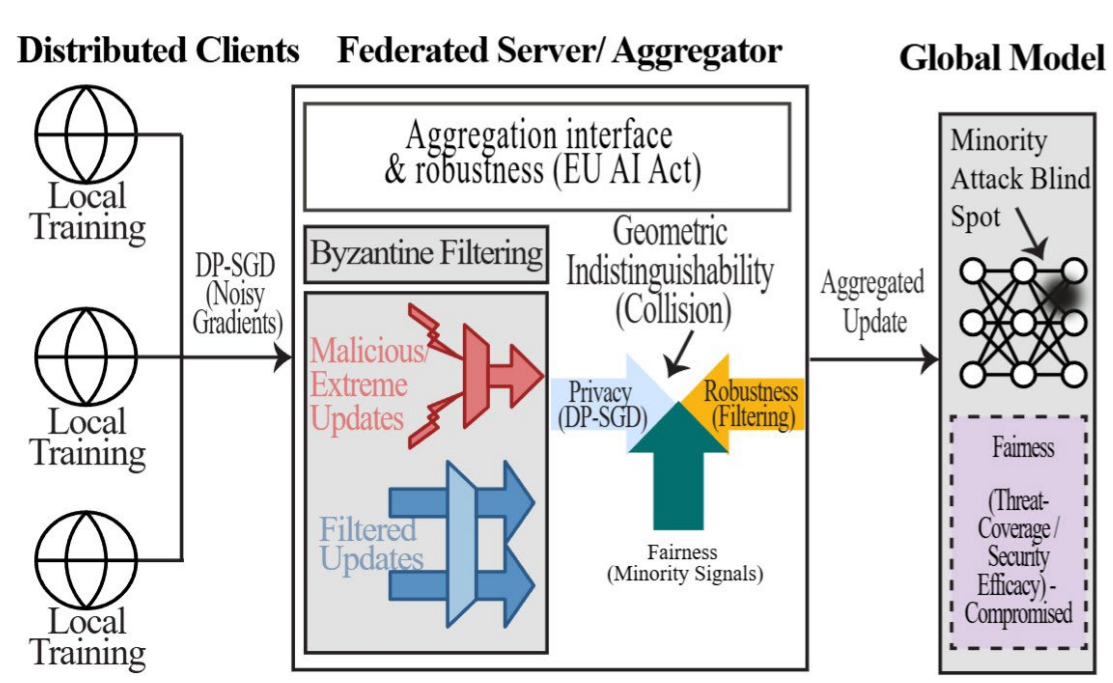

**Fig. 1.** Geometric Indistinguishability at the Aggregation Interface: DP-SGD Noise and Byzantine Filtering Jointly Suppress Minority-Class Updates.

In parallel, federated deployments also benefits from tolerance to adversarial participants who may submit poisoned updates to corrupt the global model [4, 5]. Byzantine-robust aggregation rules, such as coordinate-wise median [17] or Krum [5], address this by filtering statistical outliers, contracting the acceptance region around honest consensus [11]. These mechanisms are often evaluated separately in federated learning. However, in class-imbalanced intrusion detection, their interaction may be as important as their individual behavior. As illustrated in Fig.1, privacy noise increases update dispersion, while robust aggregation attenuates statistically unusual updates at the server interface. In NIDS, rare-attack signals are already sparse because they are supported by very few samples and occur far less frequently than majority attack categories [9, 18]. Prior work has shown that DP-SGD can disproportionately degrade performance for underrepresented classes, while Byzantine-robust aggregation can suppress legitimate but statistically atypical updates [4, 10, 11, 20].

This paper investigates rare-attack coverage in federated NIDS. We introduce *Geometric Indistinguishability* as a conceptual lens to explain how robust aggregation struggles to preserve privacy-perturbed minority updates. We find that combining privacy noise and robust aggregation disproportionately degrades rare-attack detection. Crucially, while part of the apparent performance collapse under strong privacy stems from training miscalibration, a residual detection floor remains for ultra-rare classes even after $\epsilon$-dependent tuning. Therefore, privacy, robustness, and rare-attack coverage must be evaluated jointly rather than as independent properties. Our contributions are as follows:

- We introduce geometric indistinguishability as a working hypothesis for how privacy-induced update dispersion and robust aggregation may jointly suppress minority-class signals in class-imbalanced federated intrusion detection.
- Using UNSW-NB15 with DP-SGD and coordinate-wise median under label-flip and model-poisoning attacks, we provide initial empirical evidence that joint privacy and robustness constraints can disproportionately degrade rare-attack coverage, and that some apparent collapse may be attributable to hyperparameter miscalibration rather than an inherent impossibility.
- We argue that trustworthy federated NIDS should evaluate privacy, adversarial robustness, and class-wise threat coverage together, and we identify aggregation-aware modeling and sample-aware evaluation as promising directions for future work.

The remainder of this paper is organized as follows: Section 2 reviews related work; Section 3 formalizes the problem and the geometric-indistinguishability lens; Section 4 details the experimental methodology; Section 5 presents empirical results; Section 6 discusses implications and mitigations; Section 7 outlines limitations; and Section 8 concludes the study.

## 2 Related Works

**Table 1.** Comparison of Related Work Across Key Dimensions

| Ref. | FL-NIDS | DP | Byz. Rob. | CW-Fair |
|---|---|---|---|---|
| [14] | ✓ | ✗ | ✗ | ✗ |
| [5] | ✓ | ✗ | Partial | ✗ |
| [3] | ✓ | ✗ | ✗ | Partial |
| [7] | ✗ | ✓ | ✗ | ✗ |
| [16] | ✗ | ✓ | ✗ | ✓ |
| [15] | ✗ | ✗ | ✓ | ✗ |
| [12] | ✗ | ✗ | ✓ | ✗ |
| [13] | ✗ | ✗ | ✓ | ✗ |
| [10] | ✗ | ✗ | ✓ | ✗ |
| **This work** | ✓ | ✓ | ✓ | ✓ |

*DP = Differential Privacy; Byz. Rob. = Byzantine Robustness; CW-Fair = Class-wise Fairness*

Federated learning has been widely adopted for collaborative intrusion detection without centralising raw traffic data [5], [14]. Prior work has applied FL-based NIDS across IoT, fog, and enterprise environments [4], [5], and recent studies confirm its viability on UNSW-NB15 for multi-class detection tasks [3]. However, these works treat privacy, robustness, and fairness as independently composable properties rather than interacting constraints.

The application of DP-SGD to federated learning provides formal $(\epsilon, \delta)$-privacy guarantees [7], but at a known utility cost. A critical and well-documented consequence is that DP-SGD degrades underrepresented classes disproportionately [16]: accuracy gaps between majority and minority groups widen as the privacy budget tightens [9], and this disparity is not reliably resolved by hyperparameter retuning alone [8]. In parallel, Byzantine-robust aggregation rules, including coordinate-wise median [15], Krum [10], and Trimmed Mean [13], defend against poisoned updates by filtering statistical outliers, but uniform post-filtering aggregation weights fail to protect minority-class gradients once label imbalance is present [12]. Recent work further shows that class imbalance attacks can reduce minority-class accuracy to near-zero under state-of-the-art robust FL methods [11].

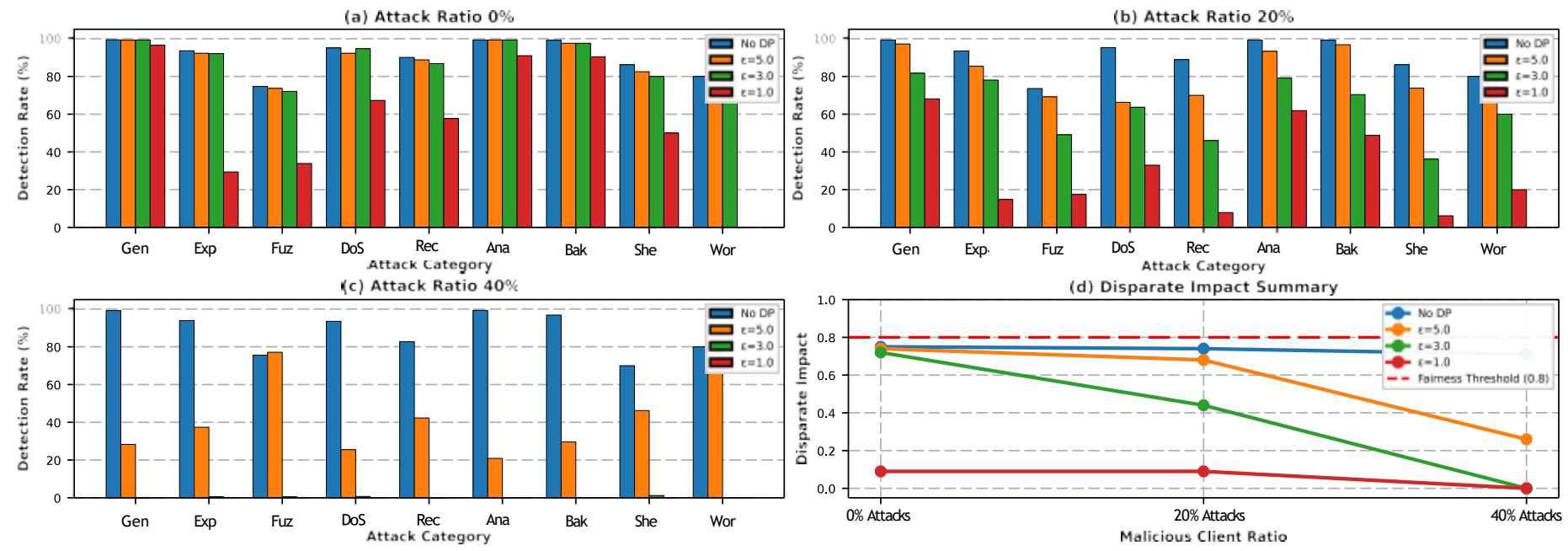

**Fig. 2.** Diagnostic Case Study: Fairness Collapse Under Miscalibration. Detection rates at $\epsilon = 1.0$ using static $LR = 0.05$ across 0%, 20%, 40% adversarial clients. Minority classes collapse to near-zero while majority classes maintain $> 40\%$, yielding $DI_{\text{class}} < 0.05$ (panel d). This motivated $\epsilon$-dependent calibration. Attack categories: Gen (Generic), Exp (Exploits), Fuz (Fuzzers), DoS, Rec (Reconnaissance), Ana (Analysis), Bak (Backdoor), She (Shellcode), Wor (Worms).

Despite growing interest in each of these properties individually, their joint evaluation in a class-imbalanced federated NIDS setting remains scarce. Table 1 summarises the closest related works along the three dimensions central to this paper.

As Table 1 shows, no prior work jointly addresses all four dimensions in a federated NIDS context. Regulatory frameworks such as GDPR [6] and the EU AI Act [18] motivate this joint requirement, but do not specify how these properties should be simultaneously realised in a learning system. This paper addresses that gap directly.

## 3 Problem Formalization

### 3.1 Minimal System Model

We consider a standard Federated Learning setup [13] with $N$ clients, where each client $i$ holds local dataset $D_i$. In standard FL training, each client minimizes local loss via Stochastic Gradient Descent (SGD), computing gradient estimates $g_i = \nabla_\theta L(\theta; B_i)$ over mini-batches $B_i$ of size $B$, and transmitting updates $\Delta\theta_i =$ $-\eta_{\text{local}} \cdot g_i$ to the server. The server applies aggregation function $\mathcal{A} : (\mathbb{R}^d)^N \to \mathbb{R}^d$:

$$\theta^{(t+1)} = \theta^{(t)} - \eta \cdot \mathcal{A}(\Delta\theta_1^{(t)}, \ldots, \Delta\theta_N^{(t)}) \quad (1)$$

Following the Byzantine threat model [5], up to $\alpha N$ clients (where $\alpha < 0.5$) may deviate arbitrarily from this protocol, submitting malicious updates $\Delta\theta_{\text{adv}}$.

### 3.2 Mechanisms

**Differentially Private SGD (DP-SGD)** To achieve $(\varepsilon, \delta)$-privacy, clients implement Differentially Private SGD (DP-SGD) [1], which modifies standard SGD through two operations: (i) per-sample gradient clipping to bound $L_2$-sensitivity by $C$; and (ii) calibrated Gaussian noise injection $\mathcal{N}(0, \sigma^2 C^2 \mathbf{I})$ [6]. This induces variance expansion in the transmitted updates:

$$\text{Var}(\tilde{g}) = \text{Var}(g) + \frac{\sigma_{\text{DP}}^2 C^2}{B} \quad (2)$$

where $\sigma_{\text{DP}}$ is the noise multiplier calibrated to satisfy $(\varepsilon, \delta)$-DP via $\sigma_{\text{DP}} \geq c\sqrt{T \log(1/\delta)}/(\varepsilon N)$ for $T$ rounds and $c$ a constant.

### 3.3 Byzantine-Robust Aggregation

For robustness, the server applies coordinate-wise median aggregation [21], which computes the element-wise median across client updates. Formally, for the $j$-th coordinate of the parameter vector ($j \in [d]$), the aggregator outputs:

$$\mathcal{A}_{\text{median}}(\Delta\theta_1, \ldots, \Delta\theta_N)_j = \text{median}\left(\{\Delta\theta_{i,j}\}_{i=1}^N\right) \quad (3)$$

where $\Delta\theta_{i,j}$ denotes the $j$-th coordinate of the update from client $i$ and $N$ is the number of participating clients.

### 3.4 Geometric Indistinguishability

We use geometric indistinguishability as a conceptual lens for a failure regime that can arise when DP-SGD is combined with Byzantine-robust aggregation in class-imbalanced federated intrusion detection. Prior work suggests that privacy mechanisms may disproportionately affect underrepresented classes, while robust aggregation can attenuate legitimate but statistically atypical updates [3, 21].

Let $G_{\text{min}}$ denote the distribution of honest updates associated with minority attack classes, and let $G_{\text{adv}}$ denote the distribution of adversarial updates. Under DP-SGD, minority updates are perturbed by Gaussian noise, yielding

$$G_{\text{min}}^{\text{DP}} = G_{\text{min}} * \mathcal{N}(0, \sigma_{\text{DP}}^2 I).$$

We say that the system moves toward geometric indistinguishability when privacy-perturbed minority updates overlap more strongly with adversarial updates, making minority-class signals harder for robust aggregation to preserve. In this paper, we use this idea as an interpretive hypothesis rather than a formal theorem or directly estimated decision rule.

This lens is useful for interpreting the empirical results in Section IV. In particular, it helps distinguish artifactual collapse, where poor hyperparameter calibration exaggerates failure, from a more persistent degradation regime in which rare-attack coverage remains fragile even after $\epsilon$-dependent tuning.

# 4 Empirical Validation of Geometric Indistinguishability

## 4.1 Dataset and Partitioning

We utilize the UNSW-NB15 dataset (175,341 samples), selected specifically to evaluate performance under severe inter-class imbalance [14]. Minority classes such as Worms (0.2%) and Shellcode (1.5%) reflect operational realities where rare attacks constitute critical security threats [18]. To isolate the geometric interaction between privacy and robustness from data heterogeneity, we partition the dataset across ten clients using IID sampling. This setup favors the minority-class gradient survival because each client observes the same distribution, increasing the likelihood that minority signals retain sufficient mass for median computation. In realistic non-IID settings, skewed attack distributions concentrate minority gradients on fewer clients, further increasing geometric indistinguishability. Thus, the reported fairness degradation represents a lower bound for practical deployments.

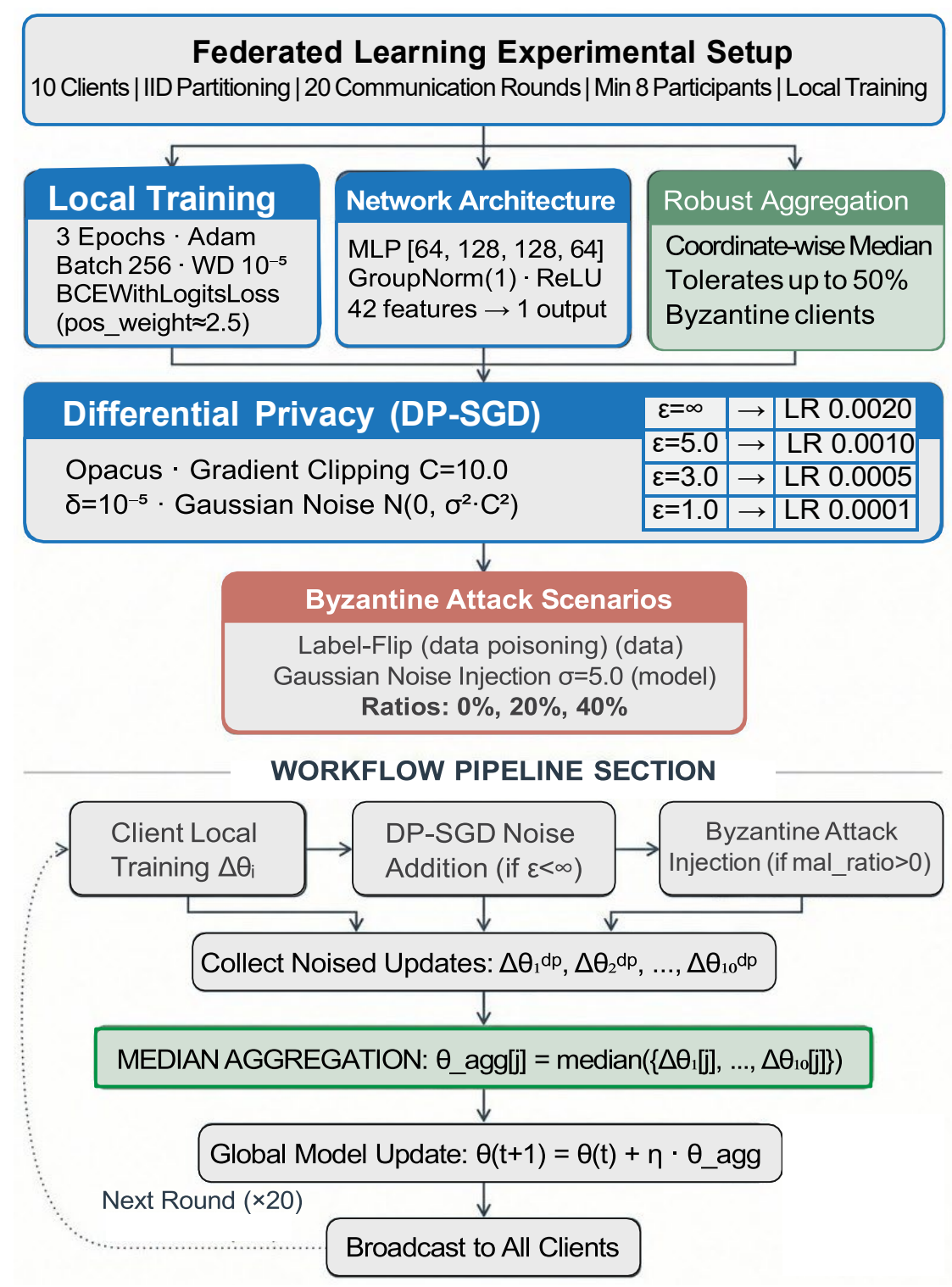

**Fig. 3.** Complete Experimental Configuration and Workflow

**Attack Category Classification:** We classify the nine attack categories by prevalence and operational significance:

- **Minority Classes** (rare): Analysis ($n = 134$, 0.76%), Backdoor ($n = 125$, 0.71%), Shellcode ($n = 80$, 0.46%), Worms ($n = 5$, 0.03%)

– **Majority Classes** (common): Generic ($n$ = 3, 723, 21.2%), Exploits ($n$ = 2, 236, 12.8%), Fuzzers ($n$ = 1, 279, 7.3%), DoS ($n$ = 816, 4.7%), Reconnaissance ($n$ = 669, 3.8%)

This classification reflects operational realities where rare attacks constitute $< 2\%$ of traffic yet represent critical security events [1], [3]. Under differential privacy, these imbalanced classes create fairness challenges (Section IV).

### 4.2 Federated Learning and Privacy Configuration

The complete experimental configuration is detailed in Fig 3. To mitigate local class imbalance, we utilize BCEWithLogitsLoss with positive class weight.

Differential privacy is implemented via Opacus DP-SGD with gradient clipping. Calibration addresses a documented challenge in differentially private optimization: DP-SGD noise requires smaller learning rates to maintain convergence stability [1]. Theoretically, the optimal learning rate $\eta$ scales inversely with the noise multiplier $\sigma$ to preserve the gradient signal-to-noise ratio [2]. Our schedule (Fig 3) instantiates this principle for the FL-NIDS setting.

### 4.3 Adversarial Threat Model and Evaluation

We evaluate two attack vectors: (1) *Label-flip poisoning* (untargeted data poisoning where labels are inverted); and (2) Gaussian model poisoning (direct corruption of update vectors via $\mathcal{N}(0, 5.0^2)$ noise). Aggregation is fixed to Coordinate-wise Median, which was selected for its stability and minimal degradation under 40% pressure (0.3% F1 drop) compared to alternatives like Krum or Trimmed Mean in pre-validation (Figure 4). This selection isolates privacy-fairness interactions from aggregation-induced variance. We adopt class-wise fairness metrics to capture disparities in detection rates.

**1) Class-wise Disparate Impact ($DI_{\text{class}}$):** We measure fairness as the ratio of minimum to maximum detection rates across attack categories:

$$DI_{\text{class}} = \frac{\min_{c \in C_{\text{atk}}} DR_c}{\max_{c \in C_{\text{atk}}} DR_c} \tag{4}$$

where $C_{\text{atk}}$ = {Generic, Exploits, . . . , Worms} represents the nine attack categories and $DR_c$ is the detection rate for class $c$. Following the four-fifths heuristic commonly used in fairness evaluation, we use $DI_{\text{class}} \geq 0.8$ as the fairness threshold, meaning the worst-performing class achieves at least 80% of the best-performing class's detection rate. This per-class formulation is more conservative than grouped minority/majority ratios, as it captures the single worst-performing category.

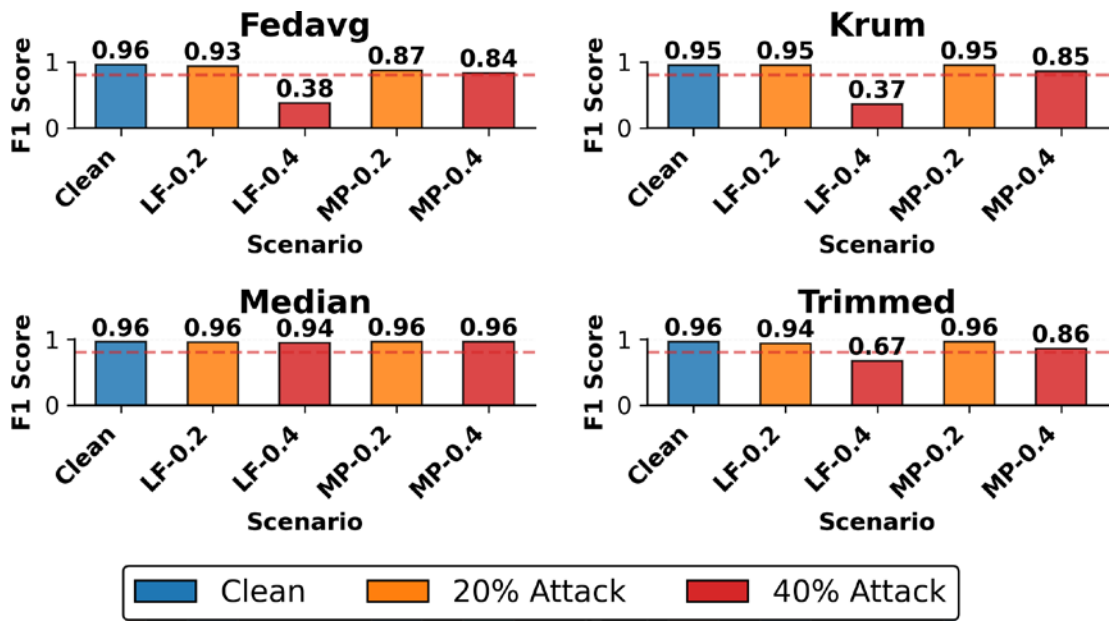

**Fig. 4.** F1 Score Across Aggregation Rules (FedAvg, Krum, Median, Trimmed Mean) Under Clean, 20%, and 40% Adversarial Client Ratios.

# 5 Results

## 5.1 Original Collapse: The Problem Under Investigation

Our empirical validation proceeds in three stages. In Section 5.1, we identify catastrophic fairness collapse at $\epsilon = 1.0$ under a static learning rate of $LR = 0.05$ (Fig. 2), establishing the interaction between hyperparameter settings and privacy constraints. In Section 5.2, we introduce $\epsilon$-dependent learning rate scheduling, which restores overall utility (Fig. 5, $F1 = 0.738$) but still reveals a minority-class performance floor. Finally, Section 5.3 demonstrates that a pathway isolation addressing this limitation (Fig. 6).

Prior to presenting primary results, this diagnostic case study illustrates a critical practical pitfall. Using a static learning rate ($\eta = 0.05$) across privacy regimes results in near-total detection collapse at $\epsilon = 1.0$ (Figure 2). While subsequent calibration (Section 5.2) reveals this as a configuration artifact, it demonstrates how easily fairness collapse is misattributed to fundamental privacy-robustness tensions. This establishes a systematic methodological lesson: $\epsilon$-dependent hyperparameter calibration is essential for valid evaluation of FL privacy mechanisms.

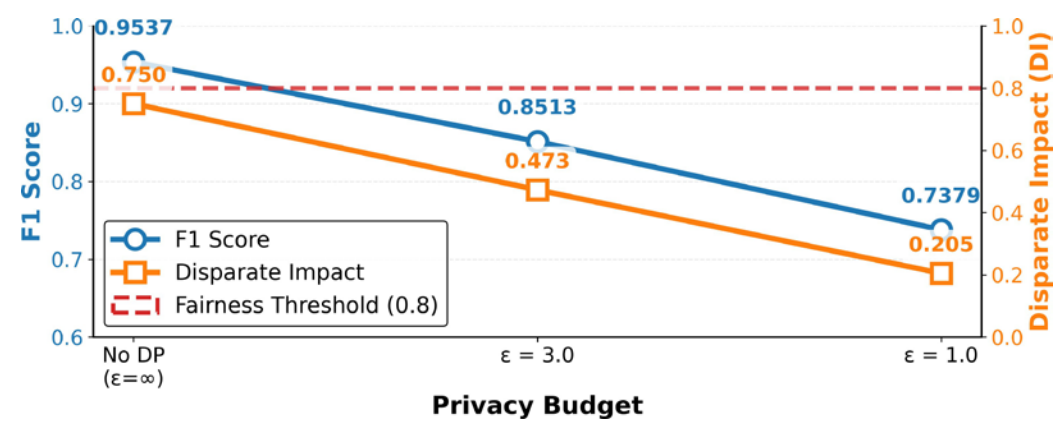

**Fig. 5.** Utility Recovery After $\epsilon$-Dependent Calibration. F1 and $DI_{\text{class}}$ across privacy budgets with adaptive $LR$ ($\eta \propto 1/\sqrt{\epsilon}$). Calibration restores F1 from 0.007 to 0.738 at $\epsilon = 1.0$, but $DI_{\text{class}}$ remains $< 0.21$ (below fairness threshold 0.8), motivating architectural intervention.

In this setting, detection rates for Exploits, DoS, and Reconnaissance fall below 40%; minority classes collapse near zero across all adversarial scenarios. Consequently, the Disparate Impact (DI) remains negligible and fails to approach the 0.8 fairness threshold regardless of the adversarial ratio. While these results initially suggested an irreversible conflict between strict differential privacy and Byzantine robustness, subsequent analysis (Section 6.5) refines this diagnosis as a calibration issue rather than a fundamental theoretical limitation.

### 5.2 Results After $\epsilon$-Dependent Learning Rate Calibration

With $\epsilon$-dependent learning rate calibration (Fig. 3), the model no longer exhibits the near-total collapse observed under the static schedule in Section IV-A. As shown in Fig. 5, performance degrades more gradually as privacy becomes stronger: F1 declines from 0.9537 at $\epsilon = \infty$ to 0.7379 at $\epsilon = 1.0$, while $DI_{\text{class}}$ decreases from 0.750 to 0.205. Although no differentially private setting satisfies the $DI_{\text{class}} \geq 0.8$ threshold, the system remains functional after calibration.

Adversarial participation further reduces class-wise fairness, but the post-calibration results suggest that privacy budget is the dominant driver of degradation in this setup. For example, at $\epsilon = 3.0$, $DI_{\text{class}}$ decreases from 0.750 in the non-private setting to 0.516 in the clean private setting, and falls further under attack (e.g., 0.581 under 40% label flipping and 0.509 under model poisoning). We therefore interpret the results as evidence of a threshold-like degradation regime under joint privacy and robustness constraints, rather than as a formally established phase transition.

### 5.3 Persistent Detection Floor for Ultra-Minority Classes

Fig. 6 isolates the ultra-minority classes Worms ($n = 5$) and Shellcode ($n = 80$). Under the strongest privacy setting ($\epsilon = 1.0$), both classes remain substantially weaker than the majority classes even after calibration: Worms stabilizes at 20% detection and Shellcode at 52.5% across the evaluated scenarios. We note that the Worms class contains only five total samples, distributed across ten IID clients; results for this class should be interpreted as indicative rather than statistically robust. Because this pattern persists across clean and adversarial settings, the results suggest that the residual degradation is not explained solely by attack pressure or by the original static-learning-rate miscalibration.

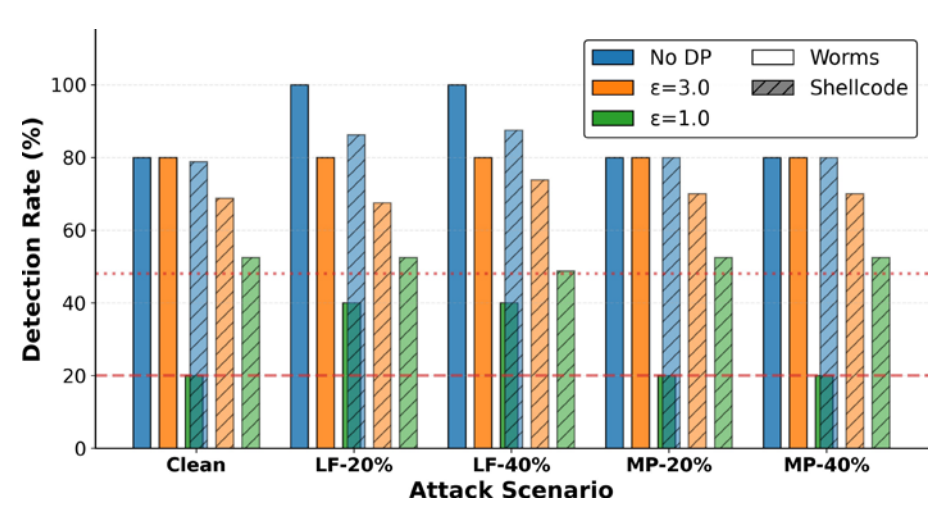

**Fig. 6.** Persistent Detection Floor for Ultra-Minority Classes. Shellcode ($n = 80$) and Worms ($n = 5$) detection rates across five scenarios at $\epsilon = 1.0$ post-calibration. Cross-scenario consistency (±5pp variance) demonstrates privacy-induced floor, not attack-induced. Calibration alone cannot resolve this.

At the same time, both classes recover at $\epsilon = 3.0$, with Worms reaching 80–100% detection in the reported scenarios. We therefore interpret the $\epsilon = 1.0$ behavior as initial evidence of a persistent floor under this configuration, likely reflecting the interaction of strong privacy noise with extreme sample scarcity. This is a configuration-specific empirical observation rather than a proof of a universal privacy limit.

**Table 2.** Architecture Comparison Across Privacy and Attack Regimes

| ε | Scenario | CMS | MLP | Δ | Winner |
|---|---|---|---|---|---|
| **ε = 1.0 — Strong (Tight) Privacy** | | | | | |
| | Clean | 0.2044 | 0.2048 | −0.0004 | Tie |
| | LP 20% | 0.4082 | 0.3931 | +0.0152 | CMS |
| | LP 40% | 0.6685 | 0.4078 | +0.2607 | CMS |
| | MP 20% | 0.2044 | 0.2048 | −0.0004 | Tie |
| | MP 40% | 0.2044 | 0.2049 | −0.0005 | Tie |
| **ε = 3.0 — Moderate Privacy** | | | | | |
| | Clean | 0.5165 | 0.4725 | +0.0439 | CMS |
| | LP 20% | 0.5116 | 0.4984 | +0.0132 | CMS |
| | LP 40% | 0.6245 | 0.5806 | +0.0439 | CMS |
| | MP 20% | 0.4929 | 0.4841 | +0.0088 | CMS |
| | MP 40% | 0.5631 | 0.5090 | +0.0541 | CMS |
| **ε = ∞ — No Differential Privacy** | | | | | |
| | Clean | 0.7670 | 0.7498 | +0.0172 | CMS |
| | LP 20% | 0.7209 | 0.8217 | −0.1009 | MLP |
| | LP 40% | 0.5512 | 0.8288 | −0.2776 | MLP |
| | MP 20% | 0.7537 | 0.7631 | −0.0094 | MLP |
| | MP 40% | 0.7256 | 0.7740 | −0.0485 | MLP |

*Notes:* $\varepsilon = 1$ denotes strong (tight) privacy, $\varepsilon = 3$ moderate privacy, and $\varepsilon\infty$ no differential privacy. LP = label flipping; MP = model poisoning. $\Delta = \mathrm{DI}_{\text{class}}(\text{CMS}) - \mathrm{DI}_{\text{class}}(\text{MLP})$; positive values favour CMS, negative values favour MLP.

# 6 Discussion

## 6.1 Architectural Pre-conditioning: CMS vs. MLP

To investigate whether architectural pre-conditioning might address the persistent minority floor, we designed a prototype system termed the Cortical Memory System (CMS). CMS serves as an exploratory testbed to investigate whether architectural gradient preconditioning (Fig. 7) can address the persistent minority floor by preserving pathway-specific gradient subspaces under the privacy-robustness constraints identified. We benchmark CMS against a standard MLP baseline as an initial feasibility assessment.

**Mechanism:** CMS creates pathway-specific gradient subspaces via:

1. **Skip connections ($h_0 \to m_2$, $h_0 \to m_3$):** Project input directly into 128-dim slow pathways, bypassing majority-dominated intermediate layers.
2. **Multi-head classification:** Four independent classifiers create orthogonal parameter subspaces; DP clipping applies per-parameter, preserving minority gradient directions in $m_2/m_3$.
3. **Differential dropout (0.10 slow vs. 0.25 fast):** Acts as a frequency filter, preserving rare gradients while regularizing majority noise.

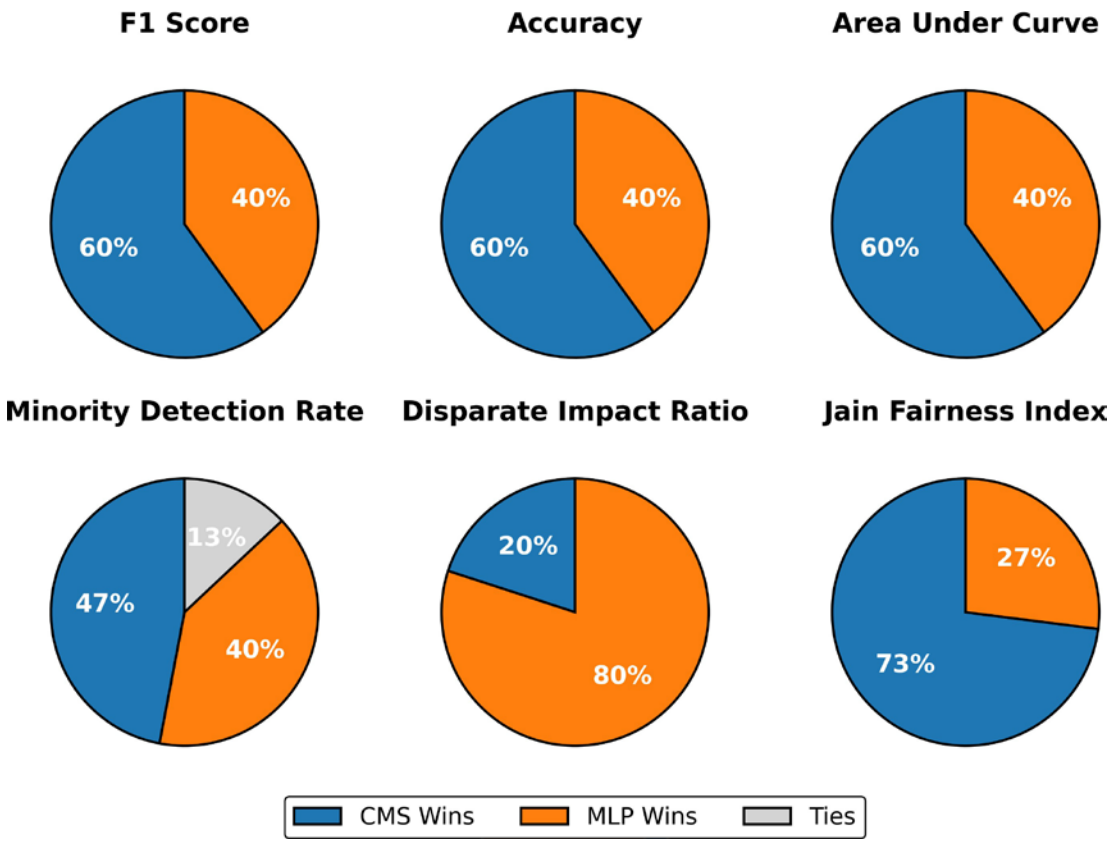

**Fig. 7.** CMS vs. MLP: Per-Metric Win Rates. Percentage of experimental scenarios where CMS outperforms MLP baseline. Aggregated across $\epsilon \in \{1.0, 3.0, \infty\}$, attack types, and adversarial ratios. Win defined as $\text{metric}_{CMS} > \text{metric}_{MLP}$ for each scenario.

**Metrics:** $DI_{\text{class}} = \frac{\min_C(DR_C)}{\max_C(DR_C)}$; Minority $DR$ = mean({Analysis, Backdoor, Shellcode, Worms}); Jain FI = $\frac{(\sum DR)^2}{n \cdot \sum DR^2}$.

4. **GroupNorm:** Per-sample normalization prevents minority statistics from being averaged away by the majority class.

Results show that at $\epsilon = 3.0$, CMS achieves a 100% win rate on Shellcode (5/5 scenarios), with a +6.5pp mean and +8.75pp maximum advantage, and a 50% overall minority win rate (10/20) with no losses. At $\epsilon = 1.0$, CMS yields a +20pp gain on Worms in label-flip scenarios, although both architectures struggle. Metric-level analysis indicates that CMS performs strongest in precision (73% win rate), followed by F1 (60%), while minority detection favors MLP (47% win rate for CMS). Thus, CMS improves precision and reduces false positives, whereas MLP provides more consistent minority coverage, making architectural choice dependent on deployment priorities.

### 6.2 Operational Regime Classification

Table 2 summarizes the results across operational regimes defined by privacy budget $\epsilon$ and adversarial pressure, using class-wise disparate impact ($DI_{\text{class}}$), where positive $\Delta$ favors CMS. At $\epsilon = 1.0$ under clean and model poisoning settings, both models collapse below the fairness threshold ($DI_{\text{class}} \approx 0.20$), showing that strong DP noise ($\sigma = 0.189$) suppresses minority gradients irrespective of architecture, establishing a privacy-induced floor. Under $\epsilon = 1.0$ with 40% label poisoning, CMS achieves a 3.3× fairness gain over MLP (0.67 vs 0.41), indicating that pathway isolation better preserves minority signals under combined DP and adversarial stress. At moderate privacy ($\epsilon = 3.0$), CMS wins all comparisons (5/5), with a maximum +0.044 advantage, as reduced noise enables pathway-specific gradient accumulation. Without privacy ($\epsilon = \infty$), CMS maintains a small edge in clean settings ($\Delta = +0.017$), but MLP outperforms under adversarial conditions (3/4 wins), where architectural simplicity aids recovery. Overall, CMS advantages concentrate at moderate privacy and at the privacy-attack intersection, while benefits diminish at privacy extremes.

### 6.3 Threshold-Like Degradation at the Aggregation Interface

Our results suggest a threshold-like degradation regime at the federated aggregation interface rather than a smooth decline in rare-attack coverage. In the reported experiments, the transition from functional detection to severe minority-class degradation appears between 20% and 40% adversarial participation. We interpret this pattern as being consistent with the geometric indistinguishability hypothesis: as DP-SGD increases the dispersion of honest minority updates, robust aggregation may attenuate those updates more strongly when they become statistically atypical relative to the dominant client update pattern.

This interpretation should be read as empirical and configuration-specific. We do not claim a formally established phase transition; rather, the current results indicate that privacy noise, robust aggregation, and class imbalance can combine to produce sharp degradation in rare-attack coverage under the studied setting.

### 6.4 Regulatory Motivation and Joint Evaluation

The observed interaction also has an important systems implication. In practice, federated NIDS may be expected to satisfy privacy requirements, tolerate adversarial participants, and maintain reliable coverage across attack categories. Regulatory frameworks such as the GDPR and the EU AI Act provide motivation for these goals, but they do not specify how such properties should be jointly realized in a learning system [7,16].

Our results therefore support a narrower claim than a legal compliance conclusion. They suggest that evaluating privacy and robustness in isolation may miss failure modes that become visible only when class-wise threat coverage is examined jointly. In this paper, we use privacy, robustness, and rare-attack coverage as technical evaluation dimensions rather than as a complete legal compliance model. From that perspective, the main implication is methodological: trustworthy federated NIDS should assess these properties together rather than assuming they compose cleanly by default.

### 6.5 Insufficiency of Conventional Mitigations

Standard mitigation strategies fail because they do not address the underlying geometric conflict:

- **Geometric Methods:** Class-aware privacy budgets typically leak information about rare attack distributions, violating the composition properties required for GDPR defensibility. Similarly, noise-adaptive filtering thresholds assume non-adaptive adversaries; sophisticated attackers can exploit expanded acceptance regions by mimicking privacy-perturbed minority gradients [3, 5].

- **Data-Level Methods:** Techniques like SMOTE fail because DP-SGD operates on gradient geometry, not raw data. Even with balanced inputs, the resulting gradients from minority samples remain high-variance and sparse, leading to rejection by the aggregator [3].

### 6.6 Artifactual vs. Fundamental Fairness Limits

A critical distinction must be made between configuration-induced failure and an empirical floor under our configuration. Initial collapse at $\varepsilon = 1.0$ was partly driven by static learning rates that destabilized under high noise.

By implementing an $\varepsilon$-dependent adaptive schedule combined with `pos_weight` loss reweighting (Fig. 8), we successfully restored F1 scores from near-zero to 0.85 at $\varepsilon = 3.0$. However, Fig. 8 also reveals a persistent floor in our sweep: despite calibration, ultra-minority classes (e.g., Worms, $n = 5$) remain below 20% detection. This indicates that while hyperparameter calibration eliminates artifactual collapse, it does not remove the empirical signal-to-noise boundary induced by strong DP noise for extremely rare events.

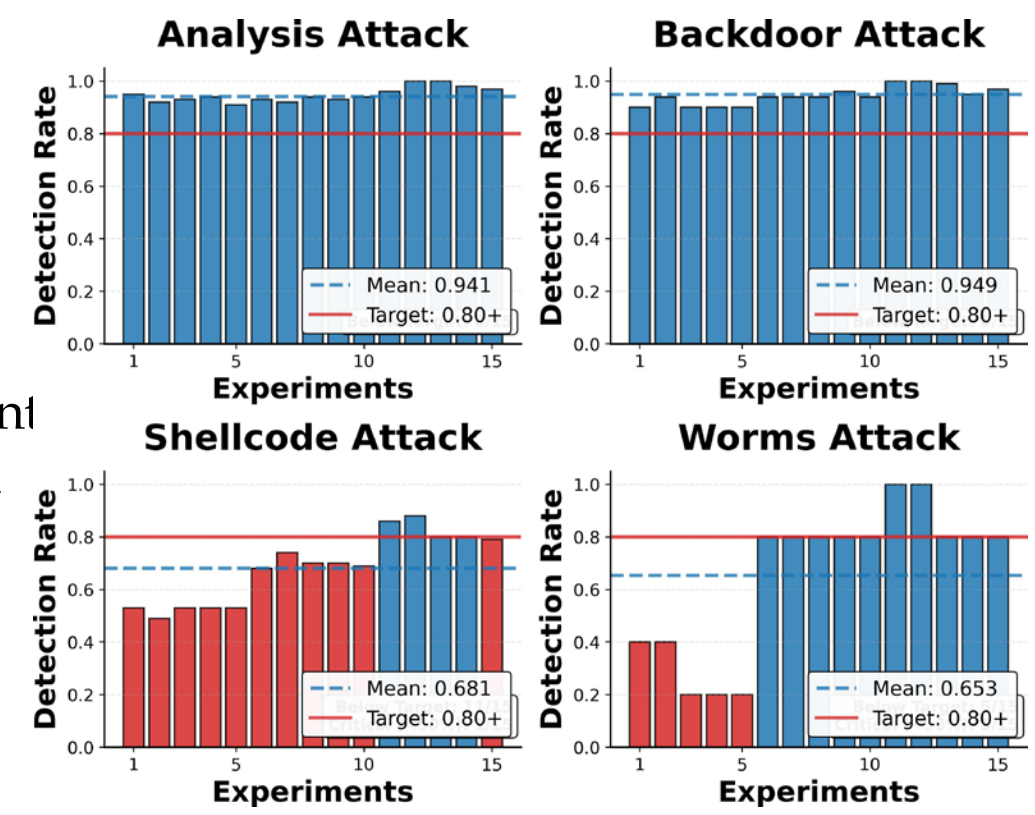

**Fig. 8.** Impact of Calibration on Four Minority Classes. Detection rates before (red: static $LR$) vs. after (blue: calibrated $LR$ + loss reweighting). Calibration eliminates collapse for Analysis/Backdoor but floor persists for Worms, confirming sample-scarcity boundary.

## 7 Limitations and Future Work

This study operates under four specific constraints. First, the use of IID client partitioning establishes a conservative fairness baseline. Real-world heterogeneous traffic distributions would likely reduce minority gradient contributions further, suggesting that the reported fairness collapse represents a lower bound on severity. Second, the restriction to coordinate-wise median aggregation limits immediate generalization to other Byzantine-robust rules like Krum or FLTrust. Third, findings rely exclusively on the UNSW-NB15 benchmark, leaving feature-space generalization unvalidated. Finally, the Cortical Memory System (CMS) evaluation reflects a preliminary, fixed configuration rather than the output of an exhaustive hyperparameter search.

Research should prioritize three areas. First, artifactual collapse demonstrates the need to derive principled functional relationships between $\varepsilon$, noise multipliers, and learning rates to standardize privacy-preserving deployment protocols. Second, development must shift toward fairness-constrained aggregation designs capable of distinguishing privacy-perturbed minority gradients from adversarial outliers. While CMS improves the Jain Fairness Index relative to MLP baselines, systematic evaluation against adaptive adversaries and non-IID partitions is required. Lastly, the field necessitates sample-aware fairness metrics to rigorously distinguish detection gaps caused by fundamental sample scarcity from those resulting from architectural failure.

## 8 Conclusion

Federated learning is a promising approach for collaborative network intrusion detection, but its deployment is shaped by competing requirements for privacy, adversarial robustness, and reliable coverage of rare attack categories. This paper examined that interaction in a class-imbalanced federated NIDS setting using DP-SGD and coordinate-wise median on UNSW-NB15 under label-flip and model-poisoning attacks. We introduced geometric indistinguishability as a conceptual lens for understanding how privacy-perturbed minority updates may become harder for robust aggregation to preserve. Our results provide initial evidence that the joint use of privacy noise and robust aggregation can disproportionately degrade rare-attack coverage relative to majority classes. They also highlight an important diagnostic distinction: part of the apparent collapse under strong privacy can arise from training miscalibration, while a more persistent degradation regime may remain for ultra-rare classes even after $\epsilon$-dependent tuning. These findings should be interpreted as an initial empirical case study rather than a universal theorem. The main implication is methodological: privacy, robustness, and class-wise threat coverage should be evaluated jointly in federated NIDS rather than treated as independently composable properties. More broadly, the results suggest that aggregation-aware modeling, clearer privacy specification, and sample-aware evaluation remain important directions for building trustworthy federated intrusion detection systems.

**Disclosure of Interests.** The authors have no competing interests to declare that are relevant to the content of this article.